\documentclass[twoside]{article}

\catcode`\@=11
\long\def\@makefntext#1{
\protect\noindent \hbox to 3.2pt {\hskip-.9pt  
$^{{\eightrm\@thefnmark}}$\hfil}#1\hfill}		

\def\@makefnmark{\hbox to 0pt{$^{\@thefnmark}$\hss}}	
	
\def\ps@myheadings{\let\@mkboth\@gobbletwo
\def\@oddhead{\hbox{}
\rightmark\hfil\eightrm\thepage}   
\def\@oddfoot{}\def\@evenhead{\eightrm\thepage\hfil
\leftmark\hbox{}}\def\@evenfoot{}
\def\sectionmark##1{}\def\subsectionmark##1{}}

\oddsidemargin=\evensidemargin
\newcounter{sectionc}\newcounter{subsectionc}\newcounter{subsubsectionc}
\renewcommand{\section}[1] {\vspace{12pt}\addtocounter{sectionc}{1} 
\setcounter{subsectionc}{0}\setcounter{subsubsectionc}{0}\noindent 
	{\tenbf\thesectionc. #1}\par\vspace{5pt}}
\renewcommand{\subsection}[1] {\vspace{12pt}\addtocounter{subsectionc}{1} 
	\setcounter{subsubsectionc}{0}\noindent 
	{\bf\thesectionc.\thesubsectionc. {\kern1pt \bfit #1}}\par\vspace{5pt}}
\renewcommand{\subsubsection}[1] {\vspace{12pt}\addtocounter{subsubsectionc}{1}
	\noindent{\tenrm\thesectionc.\thesubsectionc.\thesubsubsectionc.
	{\kern1pt \tenit #1}}\par\vspace{5pt}}
\newcommand{\nonumsection}[1] {\vspace{12pt}\noindent{\tenbf #1}
	\par\vspace{5pt}}

\newcounter{appendixc}
\newcounter{subappendixc}[appendixc]
\newcounter{subsubappendixc}[subappendixc]
\renewcommand{\thesubappendixc}{\Alph{appendixc}.\arabic{subappendixc}}
\renewcommand{\thesubsubappendixc}
	{\Alph{appendixc}.\arabic{subappendixc}.\arabic{subsubappendixc}}

\renewcommand{\appendix}[1] {\vspace{12pt}
        \refstepcounter{appendixc}
        \setcounter{figure}{0}
        \setcounter{table}{0}
        \setcounter{lemma}{0}
        \setcounter{theorem}{0}
        \setcounter{corollary}{0}
        \setcounter{definition}{0}
        \setcounter{equation}{0}
        \renewcommand{\thefigure}{\Alph{appendixc}.\arabic{figure}}
        \renewcommand{\thetable}{\Alph{appendixc}.\arabic{table}}
        \renewcommand{\theappendixc}{\Alph{appendixc}}
        \renewcommand{\thelemma}{\Alph{appendixc}.\arabic{lemma}}
        \renewcommand{\thetheorem}{\Alph{appendixc}.\arabic{theorem}}
        \renewcommand{\thedefinition}{\Alph{appendixc}.\arabic{definition}}
        \renewcommand{\thecorollary}{\Alph{appendixc}.\arabic{corollary}}
        \renewcommand{\theequation}{\Alph{appendixc}.\arabic{equation}}
        \noindent{\tenbf Appendix \theappendixc #1}\par\vspace{5pt}}
\newcommand{\subappendix}[1] {\vspace{12pt}
        \refstepcounter{subappendixc}
        \noindent{\bf Appendix \thesubappendixc. {\kern1pt \bfit #1}}
	\par\vspace{5pt}}
\newcommand{\subsubappendix}[1] {\vspace{12pt}
        \refstepcounter{subsubappendixc}
        \noindent{\rm Appendix \thesubsubappendixc. {\kern1pt \tenit #1}}
	\par\vspace{5pt}}

\newcommand{\textlineskip}{\baselineskip=13pt}
\newcommand{\smalllineskip}{\baselineskip=10pt}

\def\eightcirc{
\begin{picture}(0,0)
\put(4.4,1.8){\circle{6.5}}
\end{picture}}
\def\eightcopyright{\eightcirc\kern2.7pt\hbox{\eightrm c}} 

\newcommand{\copyrightheading}[1]
	{\vspace*{-2.5cm}\smalllineskip{\flushleft
	{\footnotesize Modern Physics Letters A, #1}\\
	{\footnotesize $\eightcopyright$\, World Scientific Publishing
	 Company}\\
	 }}

\def\abstracts#1#2#3{{
	\centering{\begin{minipage}{4.5in}\baselineskip=10pt\footnotesize
	\parindent=0pt #1\par 
	\parindent=15pt #2\par
	\parindent=15pt #3
	\end{minipage}}\par}} 

\def\keywords#1{{
	\centering{\begin{minipage}{4.5in}\baselineskip=10pt\footnotesize
	{\footnotesize\it Keywords}\/: #1
	 \end{minipage}}\par}}

\renewenvironment{thebibliography}[1]
	{\frenchspacing
	 \ninerm\baselineskip=11pt
	 \begin{list}{\arabic{enumi}.}
        {\usecounter{enumi}\setlength{\parsep}{0pt}     
	 \setlength{\leftmargin 12.7pt}{\rightmargin 0pt} 
         \setlength{\itemsep}{0pt} \settowidth
	{\labelwidth}{#1.}\sloppy}}{\end{list}}

\newcounter{itemlistc}
\newcounter{romanlistc}
\newcounter{alphlistc}
\newcounter{arabiclistc}

\newcommand{\fcaption}[1]{
        \refstepcounter{figure}
        \setbox\@tempboxa = \hbox{\footnotesize Fig.~\thefigure. #1}
        \ifdim \wd\@tempboxa > 5in
           {\begin{center}
        \parbox{5in}{\footnotesize\smalllineskip Fig.~\thefigure. #1}
            \end{center}}
        \else
             {\begin{center}
             {\footnotesize Fig.~\thefigure. #1}
              \end{center}}
        \fi}

\newcommand{\tcaption}[1]{
        \refstepcounter{table}
        \setbox\@tempboxa = \hbox{\footnotesize Table~\thetable. #1}
        \ifdim \wd\@tempboxa > 5in
           {\begin{center}
        \parbox{5in}{\footnotesize\smalllineskip Table~\thetable. #1}
            \end{center}}
        \else
             {\begin{center}
             {\footnotesize Table~\thetable. #1}
              \end{center}}
        \fi}

\def\@citex[#1]#2{\if@filesw\immediate\write\@auxout
	{\string\citation{#2}}\fi
\def\@citea{}\@cite{\@for\@citeb:=#2\do
	{\@citea\def\@citea{,}\@ifundefined
	{b@\@citeb}{{\bf ?}\@warning
	{Citation `\@citeb' on page \thepage \space undefined}}
	{\csname b@\@citeb\endcsname}}}{#1}}

\newif\if@cghi
\def\cite{\@cghitrue\@ifnextchar [{\@tempswatrue
	\@citex}{\@tempswafalse\@citex[]}}
\def\citelow{\@cghifalse\@ifnextchar [{\@tempswatrue
	\@citex}{\@tempswafalse\@citex[]}}
\def\@cite#1#2{{$\null^{#1}$\if@tempswa\typeout
	{IJCGA warning: optional citation argument 
	ignored: `#2'} \fi}}

\def\pmb#1{\setbox0=\hbox{#1}
	\kern-.025em\copy0\kern-\wd0
	\kern.05em\copy0\kern-\wd0
	\kern-.025em\raise.0433em\box0}

\def\fnt#1#2{\footnotetext{\kern-.3em
	{$^{\mbox{\scriptsize #1}}$}{#2}}}

\def\fpage#1{\begingroup
\voffset=.3in
\thispagestyle{empty}\begin{table}[b]\centerline{\footnotesize #1}
	\end{table}\endgroup}

\def\runninghead#1#2{\pagestyle{myheadings}
\markboth{{\protect\footnotesize\it{\quad #1}}\hfill}
{\hfill{\protect\footnotesize\it{#2\quad}}}}
\font\tenrm=cmr10
\font\tenit=cmti10 
\font\tenbf=cmbx10
\font\bfit=cmbxti10 at 10pt
\font\ninerm=cmr9

\font\eightrm=cmr8

\def\qed{\hbox{${\vcenter{\vbox{			
   \hrule height 0.4pt\hbox{\vrule width 0.4pt height 6pt
   \kern5pt\vrule width 0.4pt}\hrule height 0.4pt}}}$}}

\usepackage{amsmath}
\usepackage[english]{babel}
\usepackage[dvips]{graphicx}

\newcommand{\LT}   {\left}
\newcommand{\RT}   {\right}
\newcommand{\GZ}   {{\gamma Z}}
\newcommand{\MS}   {\mathrm{\overline{MS}}}
\newcommand{\Eff}  {{\text{eff}}}
\newcommand{\GeV}  {{\text{~GeV}}}
\newcommand{\Lept} {{\text{lept}}}
\newcommand{\Up}   {{\text{up}}}
\newcommand{\Down} {{\text{down}}}

\begin{document}

\runninghead{On the numerical closeness of the effective phenomenological
  electroweak $\ldots$} {On the numerical closeness of the effective
  phenomenological electroweak $\ldots$} 

\normalsize\textlineskip
\thispagestyle{empty}
\setcounter{page}{1}

\copyrightheading{}

\vspace*{0.88truein}

\fpage{1}
\begin{center}
    \textbf{ON THE NUMERICAL CLOSENESS OF THE EFFECTIVE\\[0.035truein]
      PHENOMENOLOGICAL ELECTROWEAK MIXING ANGLE $\theta$\\[0.035truein]
      AND THE $\MS$ PARAMETER $\hat\theta$} \\[0.37truein]

    \footnotesize

    M.~MALTONI \\[0.015truein]
    \textit{Dipartimento di Fisica, Universit\`a di Ferrara, I-44100
      Ferrara, Italy} \\
    \baselineskip=10pt
    \textit{Istituto TeSRE/CNR, Via Gobetti 101, I-40129 Bologna, Italy} \\
    \baselineskip=10pt
    \textit{Sezione di Ferrara, I-44100 Ferrara, Italy} \\[10pt]

    M.~I.~VYSOTSKY \\[0.015truein]
    \textit{INFN, Sezione di Ferrara, I-44100 Ferrara, Italy} \\
    \baselineskip=10pt
    \textit{ITEP, Moscow, Russia}
\end{center}


\vspace*{0.21truein}
\abstracts{It happens that $s^2$ and $\hat s^2$ are equal with $0.1\%$
  accuracy, though they are split by radiative corrections and a natural
  estimate for their difference is $1\%$. This degeneracy occurs only for
  $m_t$ value close to $170\GeV$, so no deep physical reason can be
  attributed to it. However, another puzzle of the Standard Model, the
  degeneracy of $s_\Eff^2$ and $s^2$, is not independent of the previous one
  since a good physical reason exists for $s_\Eff^2$ and $\hat s^2$
  degeneracy. We present explicit formulas which relate these three
  angles.}{}{}

\vspace*{10pt}
\keywords{electroweak angle, radiative corrections.}

\vspace*{1pt} \textlineskip
\section{Introduction}

Nowadays, when almost all LEP~I data are analyzed and published, one can
finally tell that the Standard Model is absolutely adequate to the
experimental data. The quality of fit of LEP~I, SLC and other precision data
is characterized by the value of $\chi^2/{\text{d.o.f}} =
14.4/14$~\cite{Moriond98}, which cannot be better. What can be extracted from
precision measurements for the future in addition to the bounds on the Higgs
boson mass $m_H$ (which unfortunately are rather
weak~\cite{Moriond98,NOV98})? As everybody knows, it is the value of $\hat
s_Z^2 \equiv \sin^2 \hat\theta_Z$ which is used to study gauge couplings
unification in the framework of GUT models. The corresponding angle is
calculated in the modified minimal subtraction scheme~($\MS$), with $\mu =
m_Z$. From~\cite{Langacker96} one can see that this quantity appears to be
numerically very close to the phenomenological parameter $s^2 \equiv
\sin^2\theta$, which is defined by the best measured quantities $G_F$, $m_Z$
and $\bar\alpha \equiv \alpha(m_Z)$:
\begin{gather}
    c^2 s^2 \equiv \cos^2 \theta \sin^2 \theta = 
      \frac{\pi \bar\alpha}{\sqrt{2} G_F m_Z^2}, \\
    \label{eq10b} s^2 = 0.2311(2),
\end{gather}
and which was used to describe electroweak precision data in a natural way
(for review and references see~\cite{NOV96}). This should be compared
with~\cite{Langacker96}:\footnote{According to the 1998 edition of RPP,
$\bar{c} = 1.0003(7)$, and $\hat s_Z^2 = 0.23124(24)$.}
\begin{equation}
    \hat s_Z^2 = \bar{c}(m_t, m_H) s^2 = 1.002(1) \cdot s^2 = 0.2316(2).
\end{equation}

The aim of the present paper is to present a formula which provides the
relation between $s^2$ and $\hat s^2 \equiv \sin^2 \hat\theta$. Analyzing it
we will see that this numerical coincidence occurs only for the top quark
mass $m_t$ close to $170\GeV$, so it is really a coincidence without any
physical explanation. At this point it is useful to remind that there is one
more coincidence in the Standard Model: $s_\Eff^2 \equiv \sin^2 \theta_\Eff$,
which describes asymmetries in $Z$ boson decays, happens to be very close to
$s^2$. And also this occurs only for $m_t$ close to $170\GeV$. However,
writing the expression for $\hat s^2$ through $s_\Eff^2$ we will see that
these two angles are naturally close, and their coincidence does not depend
on the top mass and has a straightforward physical explanation. In this way
we will see that, instead of two accidental coincidences between three
mixing angles, we have only one.

\section{$\hat s^2$ versus $s^2$}

To get necessary formulas we should start from the expression for the $\MS$
quantity $\hat s^2$. By definition, 
\begin{gather}
    \label{eq30a} \hat c = \frac{\hat g_0}{\hat{\bar g}_0}, \\
    \label{eq30b} \hat c^2 + \hat s^2 = 1,
\end{gather}
where $\hat g_0$ and $\hat{\bar g}_0$ are $W$ and $Z$ boson bare coupling
constants defined in $\MS$ renormalization scheme with $\mu = m_Z$. The
simplest way to get the expression for $\hat s^2$ in terms of $s^2$ and the
combination of polarization operators is to follow the procedure discussed
in~\cite{NOV93}. That is, to write the expressions for $G_F$, $m_Z$ and
$\bar\alpha$ through bare parameters plus radiative corrections and to solve
them for bare charges through $\cos\theta$, $\sin\theta$ and radiative
corrections. At a certain stage, angle $\theta_0$ was introduced
in~\cite{NOV93} ($c_0 \equiv \cos\theta_0 \equiv g_0 / \bar g_0 = m_{W0} /
m_{Z0}$), and the following expression for its cosine was obtained:
\begin{equation} \label{eq50}
    c_0 = c - \frac{c s^2}{2 \LT( c^2 - s^2 \RT)} 
    \LT( 2 \frac{s}{c} \Pi_\GZ(0) + \Pi_\gamma(m_Z^2) - \Pi_Z(m_Z^2)
    + \Pi_W(0) + D \RT),
\end{equation}
where $D$ comes from the box and vertex radiative corrections to muon decay,
and $\Pi_i$ are the polarization operators. This angle $\theta_0$ will
coincide with $\hat\theta$ if $D$ and $\Pi_i$ are calculated in $\MS$
framework with $\mu = m_Z$. From~\eqref{eq50} we easily get:
\begin{equation} \label{eq70}
    \hat s^2 = s^2 + \frac{c^2 s^2}{c^2 - s^2}
    \LT( 2 \frac{s}{c} \hat\Pi_\GZ(0) + \hat\Pi_\gamma(m_Z^2) 
    - \hat\Pi_Z(m_Z^2) + \hat\Pi_W(0) + \hat D \RT),
\end{equation}
where the quantities with a hat are calculated in $\MS$. Since the last
equation is central for the present paper, let us give a different
derivation of it. We start from the formulas for the vector boson masses
which take place in $\MS$ renormalization scheme:
\begin{gather}
    \label{eq80a} m_W^2 = \frac{\hat g_0^2 \hat\eta_0^2}{4}
      - \hat\Sigma_W(m_W^2), \\
    \label{eq80b} m_Z^2 = \frac{\hat{\bar g}_0^2 \hat\eta_0^2}{4}
      - \hat\Sigma_Z(m_Z^2), 
\end{gather}
where $\Sigma_i(q^2) \equiv \Pi_i(q^2) m_i^2$. From eqs.~\eqref{eq30a},
\eqref{eq30b}, \eqref{eq80a} and~\eqref{eq80b} we get (see
also~\cite{Sirlin89}):
\begin{equation} \begin{split} \label{eq100}
    \hat s^2 = 1 - \frac{\hat g_0^2}{\hat{\bar g}_0^2}
      & = 1 - \frac{m_W^2 + \hat\Sigma_W(m_W^2)}
        {m_Z^2 + \hat\Sigma_Z(m_Z^2)} \\
      & = 1 - \frac{m_W^2}{m_Z^2} + c^2
      \LT[ \hat\Pi_Z(m_Z^2) - \hat\Pi_W(m_W^2) \RT],
\end{split} \end{equation}
where in the last expression we have substituted $\LT( m_W/m_Z \RT)^2$ with
$c^2$ in the factor which multiplies $\Pi_i$, which is correct at one-loop
level. Now for the ratio $m_W/m_Z$ in the last expression in~\eqref{eq100}
we should use a formula which takes radiative corrections into account. We
follow a general approach to the electroweak radiative corrections presented
partly in~\cite{NOV93}, so we use eq.~(38) from that paper:
\begin{equation} \begin{split}
    \frac{m_W^2}{m_Z^2} = c^2 & + \frac{c^2 s^2}{c^2 - s^2}
    \bigg( \frac{c^2}{s^2} \LT[ \Pi_Z(m_Z^2) - \Pi_W(m_W^2) \RT] \\
    & + \Pi_W(m_W^2) - \Pi_W(0) - \Pi_\gamma(m_Z^2)
    - 2 \frac{s}{c} \Pi_\GZ(0) - D \bigg).
\end{split} \end{equation}

\begin{figure}[!h] \centering
    \includegraphics[width=12cm,height=8cm]{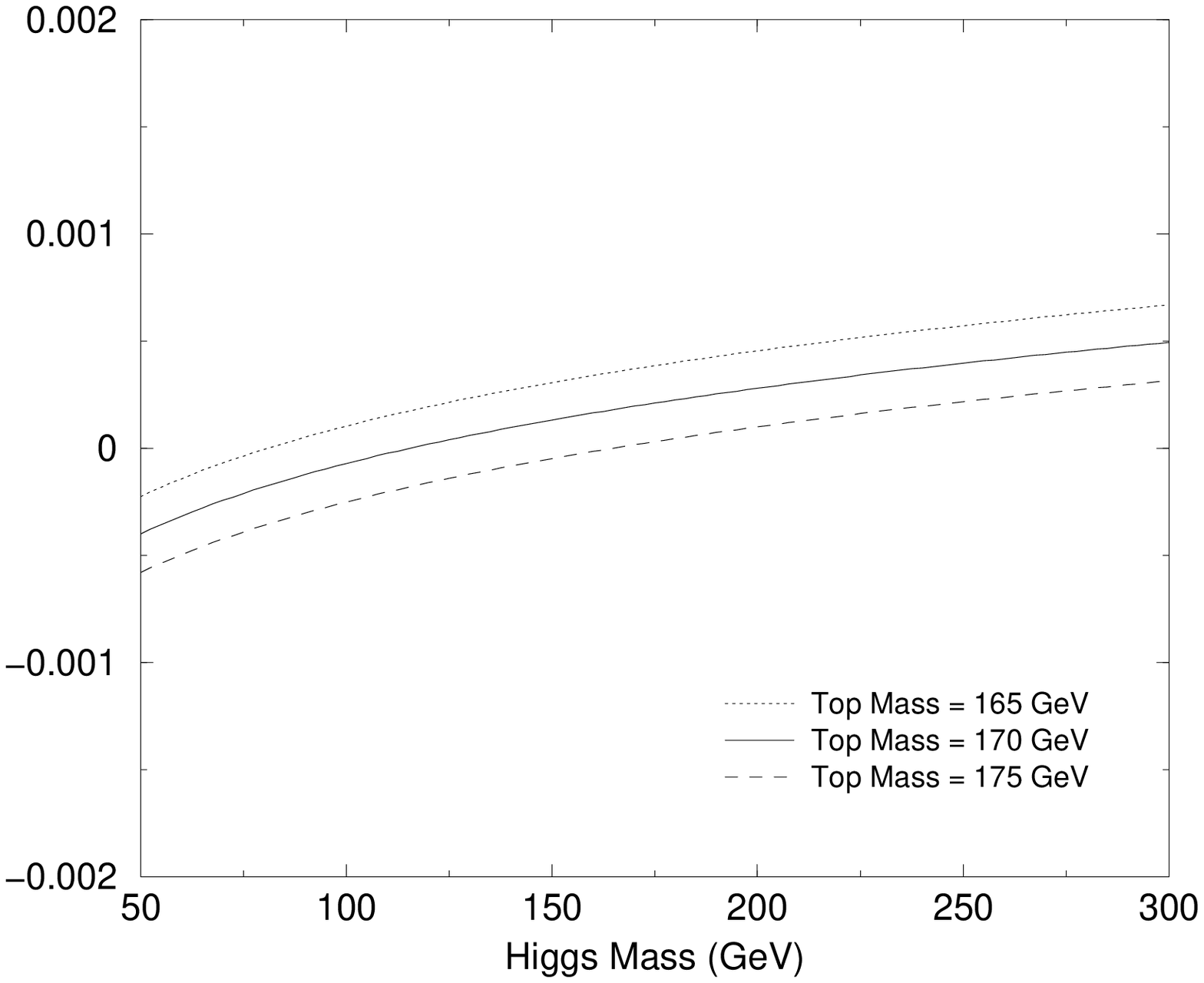} \\[5pt]
    \includegraphics[width=12cm,height=8cm]{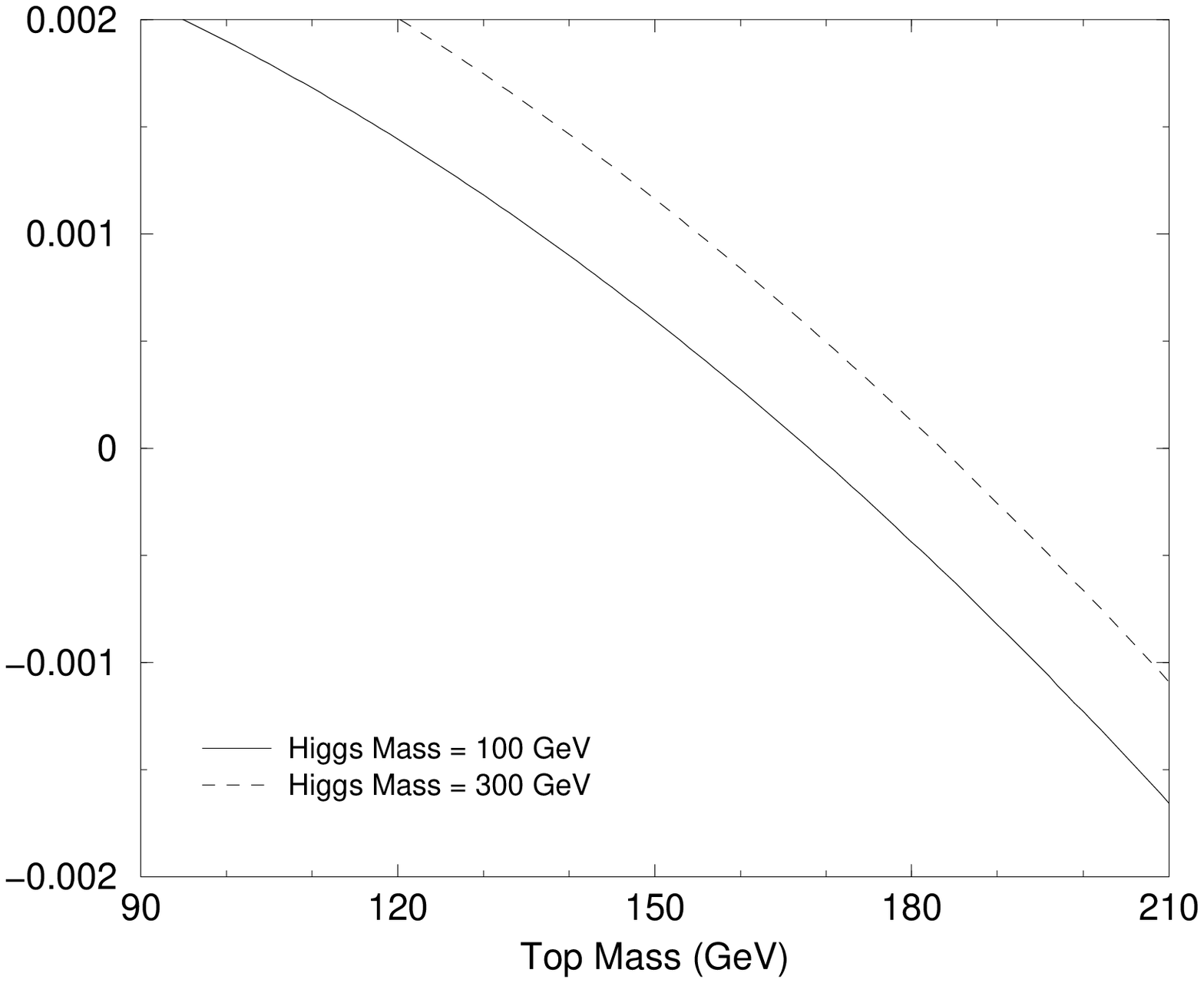}
    \caption{$\hat s^2 - s^2$ as a function of the Higgs mass $m_H$ and of
      the top mass $m_t$.}
    \label{fig10}
\end{figure}

Since both $m_W/m_Z$ and $c$ are finite, the expression for the radiative
corrections is finite as well and we can use $\MS$ quantities in it:
\begin{equation} \begin{split}
    \frac{m_W^2}{m_Z^2} = c^2 & + \frac{c^2 s^2}{c^2 - s^2} \bigg(
    \frac{c^2}{s^2} \LT[ \hat\Pi_Z(m_Z^2) - \hat\Pi_W(m_W^2) \RT] \\
    & + \hat\Pi_W(m_W^2) - \hat\Pi_W(0) - \hat\Pi_\gamma(m_Z^2)
    - 2 \frac{s}{c} \hat\Pi_\GZ(0) - \hat D \bigg).
\end{split} \end{equation}
Substituting the last equation in~\eqref{eq100}, we obtain:
\begin{equation} \label{eq130}
    \hat s^2 = s^2 + \frac{c^2 s^2}{c^2 - s^2} 
    \LT( 2 \frac{s}{c} \hat\Pi_\GZ(0) + \hat\Pi_\gamma(m_Z^2)
    - \hat\Pi_Z(m_Z^2) + \hat\Pi_W(0) + \hat D \RT),
\end{equation}
which coincides with eq.~\eqref{eq70}. In figure~\ref{fig10} we show the
$\hat s^2 - s^2$ dependence on $m_H$ and $m_t$.\footnote{In order to take
into account top and $W$ boson contributions to $\alpha$, we used $s^2 +
0.00015$ instead of $s^2$ in~\eqref{eq130} (see~\cite{NOV94a}).} It is clear
that $\hat s^2$ is close to $s^2$ only for $m_t$ around $170\GeV$, so one
cannot find any physical reason for the closeness of these two angles. The
fact that $\hat s^2 - s^2$ rapidly varies with $m_t$ can be figured out from
the large $m_t$ approximation:
\begin{equation}
    \LT. \hat s^2 - s^2 \RT|_{m_t \gg m_Z} \approx
    -\frac{3 \bar\alpha}{16 \pi \LT( c^2 - s^2 \RT)}
    \LT( \frac{m_t}{m_Z} \RT)^2.
\end{equation}

At this point we state that the numerical closeness of $\hat s^2$ and $s^2$
is a mere coincidence without any deep physical reason. However, the reason
exists for the closeness of $\hat\theta$ and another electroweak mixing
angle, $\theta_\Eff$. On the other hand, $\theta_\Eff$ appeared to be
numerically close to $\theta$ and this solves the puzzle (according to the
last data fit, $\sin^2\theta_\Eff^\Lept = 0.2315(2)$).

\section{$\hat s^2$ versus $s_\Eff^2$}

The quantity $s_\Eff$ describes the asymmetries in $Z$ boson decay;
$s_\Eff^\Lept$, $s_\Eff^\Up$ and $s_\Eff^\Down$ refers to decays into pairs
of leptons, up-quarks and down-quarks, respectively. Let us discuss
$s_\Eff^\Lept$. Using eq.~(73) from~\cite{NOV93}, we easily obtain:
\begin{equation} \begin{split} \label{eq180}
    {s_\Eff^\Lept}^2 = s^2 & + cs \LT[ F_V^{Ze}
    - \LT( 1 - 4 s^2 \RT) F_A^{Ze} \RT] + cs \Pi_\GZ(m_Z^2) \\
    & + \frac{c^2 s^2}{c^2 - s^2} \LT[ \Pi_\gamma(m_Z^2) - \Pi_Z(m_Z^2)
    + \Pi_W(0) + 2 \frac{s}{c} \Pi_\GZ(0) + D \RT],
\end{split} \end{equation}
where $F_A^{Ze}$ and $F_V^{Ze}$ describe the radiative corrections to $Zee$
axial and vector vertices.

Since both $s_\Eff^2$ and $s^2$ are finite, equality~\eqref{eq180} will be
correct if all radiative corrections are calculated in $\MS$ scheme as well.
Comparing equations~\eqref{eq70} and~\eqref{eq180} we get (see
also~\cite{Gambino94}):
\begin{equation} \label{eq190}
    \hat s^2 = {s_\Eff^\Lept}^2 - cs \LT[ \hat F_V^{Ze} 
    - \LT( 1 - 4 s^2 \RT) \hat F_A^{Ze} \RT] - cs \hat\Pi_\GZ(m_Z^2).
\end{equation}
The form of the last equation can be foreseen without any calculation. The
point is that both $\hat\theta$ and $\theta_\Eff$ are defined by the ratio
of bare gauge coupling constants; the difference between them arises since
$\theta_\Eff$ describes $Z \to e^+ e^-$ decays and in this case the
additional vertex radiative corrections as well as $Z \to \gamma \to e^+
e^-$ transition contribute to $\theta_\Eff$. In~\eqref{eq190} these
additional terms are subtracted from ${s_\Eff^\Lept}^2$ in order to get
$\hat s^2$. The vertex term in~\eqref{eq190} is a mere number, while
$\hat\Pi_\GZ$ depends on $m_t$ only logarithmically due to the
non-decoupling property of $\MS$ scheme (since a diagonal vector current is
conserved, there is no $m_t^2$ term in $\hat\Pi_\GZ$, that is why $\hat
\Pi_\GZ$ is numerically small). There is also no $m_H$ dependence in the
difference $\hat s^2 - s_\Eff^2$.

From the $\hat\Pi_\GZ(m_Z^2)$ term we get the following expression for the
logarithmically enhanced contribution for $m_t \gg m_Z$:
\begin{equation} \label{eq195}
    \LT. \hat s^2 - {s_\Eff^\Lept}^2 \RT|_{m_t \gg m_Z} \approx
    \frac{\bar\alpha}{\pi} \LT( \frac16 - \frac49 s^2 \RT)
    \ln \LT( \frac{m_t}{m_Z} \RT)^2.
\end{equation}

Having all the necessary formulas in our disposal, we are ready to make
numerical estimates. Using expressions~(93),~(94) from~\cite{NOV96} and
formulas from Appendix G of~\cite{NOV93}, we get:
\begin{align}
    \label{eq200a} \hat F_V^{Ze} = 0.00197 + \frac{\bar\alpha}{8 \pi}
      \frac{c}{s^3} \ln \LT( \frac{m_W}{m_Z} \RT)^2 = 0.00133, \\
    \label{eq200b} \hat F_A^{Ze} = 0.00186 + \frac{\bar\alpha}{8 \pi}
      \frac{c}{s^3} \ln \LT( \frac{m_W}{m_Z} \RT)^2 = 0.00122,
\end{align}
where the logarithmic terms arise from the divergent parts of vertex
functions after imposing $\MS$ renormalization conditions with $\mu = m_Z$.
Note that in numerical calculations we substituted $c^2$ for
$\LT( m_W/m_Z \RT)^2$.

To calculate $\hat\Pi_\GZ(m_Z^2)$ we use formulas from Appendices of
paper~\cite{NOV93}, which take into account $W^+ W^-$, light fermions and
$(t,b)$ doublet contributions. For $m_t = 170\GeV$ we obtain:
\begin{equation} \label{eq210}
    \hat\Pi_\GZ(m_Z^2) = -0.00119.
\end{equation}

Substituting~\eqref{eq200a},~\eqref{eq200b} and~\eqref{eq210}
into~\eqref{eq190}, we finally obtain:
\begin{equation} \label{eq220}
    \hat s^2 = {s_\Eff^\Lept}^2 - 0.00052 + 0.00050
    = {s_\Eff^\Lept}^2 - 0.00002,
\end{equation}
where an accidental cancellation between vertex and $\hat\Pi_\GZ$
contributions occurs (see also~\cite{Gambino94}). This cancellation is
peculiar to ${s_\Eff^\Lept}^2$ and does not occur for ${s^2}_\Eff^\Up$ or
${s^2}_\Eff^\Down$. As a consequence, for ${s_\Eff^\Lept}^2$ and $\hat s^2$
difference 2~loop contributions can be comparable or even larger than 1~loop.

Now we will calculate the leading two loop corrections. They are of the
order of $\alpha \alpha_s$ and come from the insertion of a gluon into quark
loops which contribute to $\hat\Pi_\GZ(m_Z^2)$. There are two types of
one-loop diagrams: with light quarks ($u,d,c,s,b$) and with heavy top ($t$).
We extract necessary 2-loop formulas from the Kniehl paper~\cite{Kniehl90}.
However, in that article all calculations were made with ultraviolet cutoff
$\Lambda$. To convert to $\MS$ we compare these formulas with calculations
of Djouadi and Gambino~\cite{Kniehl90}. In this way we find the following
replacement rule:
\begin{equation}
    \ln \LT( \frac{\Lambda^2}{m_Z^2} \RT) \to \Delta_Z + 
    \frac{55}{12} - 4 \zeta(3) = \frac{55}{12} - 4 \zeta(3)
    \approx -0.225,
\end{equation}
where the last equality holds for $\mu = m_Z$.

For the case of light quarks contribution ($u,d,c,s,b$), we get:
\begin{equation} \label{eq240}
    \delta_{\text{light}}^{\alpha \alpha_s} \hat\Pi_\GZ (m_Z^2) =
    \frac{\hat\alpha_s(m_Z)}{\pi} \frac{\bar\alpha}{\pi cs}
    \LT(\frac{7}{12} - \frac{11}{9} s^2 \RT)
    \LT[ \frac{55}{12} - 4 \zeta(3) \RT] \approx -0.00002,
\end{equation}
where we use $\hat\alpha_s(m_Z) = 0.12$ for numerical estimate. For the
contribution of the top quark we obtain:
\begin{equation} \begin{split} \label{eq250}
    \delta_t^{\alpha \alpha_s} \hat\Pi_\GZ (m_Z^2) & =
    \frac{\hat\alpha_s(m_t)}{\pi} \frac{\bar\alpha}{\pi cs}
    \LT(\frac{1}{6} - \frac{4}{9} s^2 \RT)
    \LT[ \frac{55}{12} - 4 \zeta(3) - \ln{t}
    + 4 t V_1\LT(\frac{1}{4 t}\RT) \RT] \\
    & \approx 0.00004,
\end{split} \end{equation}
where $t \equiv (m_t/m_Z)^2$ and~\cite{NOV94b,Kniehl90}:
\begin{gather}
    \hat\alpha_s(m_t) = \frac{\hat\alpha_s(m_Z)}
      {1 + \frac{23}{12\pi} \hat\alpha_s(m_Z) \ln{t}} \approx 0.11, \\[5pt]
     \label{eq260b} V_1(x) = \LT[ 4 \zeta(3) - \frac{5}{6} \RT]x
       + \frac{328}{81} x^2 + \frac{1796}{675} x^3 + {\ldots}, \\[5pt]
    \zeta(3) = 1.2020569{\ldots} \; .
\end{gather}
Substituting~\eqref{eq240} and~\eqref{eq250} into~\eqref{eq190}, we find:
\begin{equation} \label{eq270}
    \hat s^2 = {s_\Eff^\Lept}^2 - 0.00002 - 0.00001,
\end{equation}
where the first number corresponds to the corrections of order $\alpha$
shown in~\eqref{eq220}, while the second to corrections of the order $\alpha
\alpha_s$.

Since the leading $\sim \alpha$ correction cancel almost completely
in~\eqref{eq220}, one start to worry about significance of two loop
$\alpha^2$ corrections. Enhanced $\alpha^2 t$ correction in~\eqref{eq190}
was calculated in~\cite{Gambino96}, where it is stated that it is
numerically negligible; $\alpha^2$ corrections are not calculated yet.
However, according to~\cite{Gambino96} there exist enhanced two-loop
$\alpha^2 \pi^2$ correction, which come from the interference of the
imaginary parts of $\Pi_\GZ$ and $\Pi_\gamma$. Numerically it
gives~\cite{Gambino96}:
\begin{equation} \label{eq280}
    \delta_{\text{int}}^{\alpha^2} \LT( \hat s^2 - {s_\Eff^\Lept}^2 \RT) 
    = -0.00004.
\end{equation}
Adding~\eqref{eq280} to~\eqref{eq270} we finally get:
\begin{equation} \label{eq290}
    \hat s^2 = {s_\Eff^\Lept}^2 - 0.00007.
\end{equation}
It is instructive to compare the last formula with the corresponding numbers
in Tables~1 and~2 from~\cite{Degrassi97} as well as the last formula
in~\cite{Gambino96}.

\begin{figure}[!b] \centering
    \includegraphics[width=12cm,height=8cm]{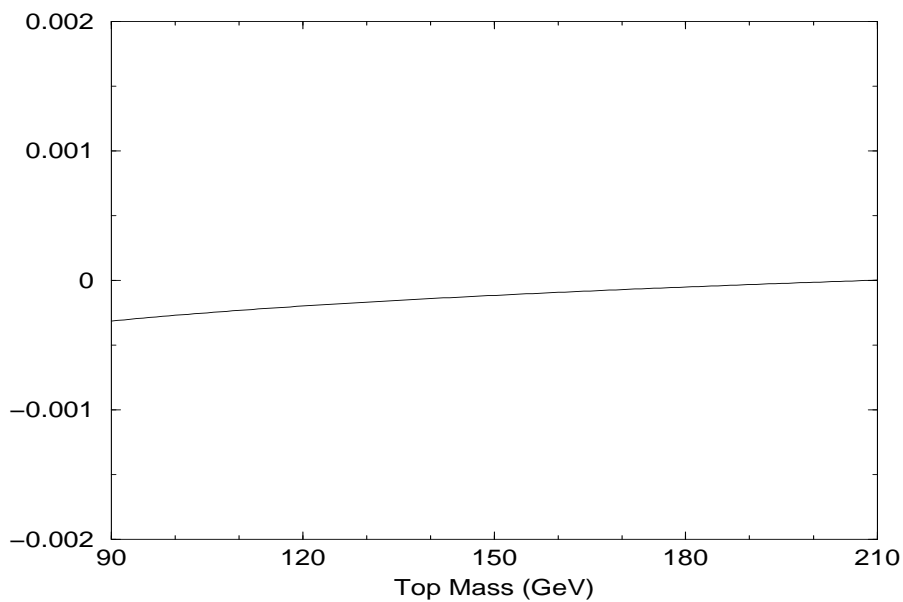}
    \caption{$\hat s^2 - {s_\Eff^\Lept}^2$ as a function of the top mass
      $m_t$. This quantity does not depend on the Higgs mass.}
    \label{fig30}
\end{figure}

In figure~\ref{fig30} the dependence of $\hat s^2 - {s_\Eff^\Lept}^2$ on
$m_t$ is presented. One can easily see that, unlike the case of $\hat s^2 -
s^2$ difference, here the dependence on $m_t$ is really small for large
$m_t$ values interval.

\section{Conclusions}

Coming back to the title of the present paper, we should study
eq.~\eqref{eq130} in more details. From this equation (or looking at
fig.~\ref{fig10}) one can see that, for $m_t=170\GeV$ and $m_H=111\GeV$,
$\hat s^2$ equals $s^2$ with high accuracy:
\begin{equation} \label{eq295}
    \LT. \hat s^2 - s^2 \RT|_{\substack{m_t = 170\GeV \\ m_H = 111\GeV}}
    = -0.00002.
\end{equation}
Taken into account ``theoretical'' prediction:
\begin{equation} \label{eq300}
    {s_\Eff^\Lept}^2 = 0.2315,
\end{equation}
which is valid for $m_t=170\GeV$, $m_H=111\GeV$, and
comparing~\eqref{eq10b},~\eqref{eq295} with~\eqref{eq290}, we observe
evident inconsistency. To overcome it small higher loop corrections
in~\eqref{eq130} should be accounted for, in analogy with what was done in
eqs.~\eqref{eq240} and~\eqref{eq250}. One can act straightforwardly, taking
into account corrections to polarization operators entering~\eqref{eq130}.
Another possible way is to take expression for $\hat s^2$ through
${s_\Eff^\Lept}^2$ (eq.~\eqref{eq190}) and to use in it expression for
${s_\Eff^\Lept}^2$ through $s^2$ and higher order radiative corrections:
\begin{equation} \begin{split} \label{eq310}
    \delta \LT( \hat s^2 - s^2 \RT) = & - cs \delta^{\alpha \alpha_s}
    \hat \Pi_\GZ(m_Z^2) \\
    & - \frac{3 \bar\alpha}{16\pi \LT( c^2 - s^2 \RT)}
    \LT( \delta_2 V_R + \delta_3 V_R + \delta_4 V_R + \delta'_4 V_R\RT),
\end{split} \end{equation}
where we take into account that in expression for ${s_\Eff^\Lept}^2$ through
$s^2$ radiative corrections are finite, so $\MS$ subtraction should not be
imposed; expressions for $\delta_i V_R$ can be found in~\cite{NOV96}
and~\cite{NOV95}:
\begin{align}
    \label{eq320a} \delta_2 V_R(t,h) & = \frac{4}{3}
      \frac{\hat\alpha_s(m_t)}{\pi} \LT[ t A_1\LT(\frac{1}{4t}\RT)
      - \frac{5}{3} t V_1\LT(\frac{1}{4t}\RT)
      - 4 t F_1(0) + \frac{1}{6} \ln t \RT], \\ 
    \delta_3 V_R(t,h) & = -14.594 \frac{\hat\alpha_s^2(m_t)}{\pi^2} t, \\
    \delta_4 V_R(t,h) & = -\frac{\bar\alpha}{16 \pi s^2 c^2}
      A\LT(\frac{m_H}{m_t}\RT) t^2, \\
    \label{eq320d} \delta'_4 V_R(t,h) & = 
      -\frac{3 \bar\alpha}{16 \pi \LT( c^2 - s^2 \RT)^2} t^2
\end{align}
where the expression for $V_1$ is given in~\eqref{eq260b} and expressions
for $A_1$, $F_1$ and $A$ can be found in~\cite{NOV96} and~\cite{NOV95}.

Substituting eqs.~\eqref{eq240},~\eqref{eq250}
and~\eqref{eq320a}-\eqref{eq320d} into~\eqref{eq310}, taking into
account eq.~\eqref{eq280} and making numerical estimate, we get:
\begin{gather}
    \delta \LT( \hat s^2 - s^2 \RT) = 0.00042, \\
    \hat s^2 = s^2 - 0.00002 + 0.00042 = 0.2315
\end{gather}
($-0.00002$ comes from~\eqref{eq295}), which is quite close to~\eqref{eq290}
(taking into account the numerical value of ${s_\Eff^\Lept}^2$
from~\eqref{eq300}).

Let us mention that in generalizations of Standard Model a lot of new heavy
particles occur, and all of them contribute to $\hat s^2$ due to the
non-decoupling property of $\MS$ renormalization. To avoid this
non-universality of $\MS$ quantities, it was suggested to subtract
contributions of the particles with masses larger than $\mu$ from
$\Pi_\gamma$ and $\Pi_\GZ$, and in particular, to subtract the logarithmic
term shown in~\eqref{eq195} from the $\hat s^2$
value~\cite{Marciano90,Langacker96}. According to the definition
accepted in~\cite{Langacker96}, the quantity $\hat s^2$ which has been
discussed up to now is called $\hat s_{\text{ND}}^2$, while a new
``decoupled'' $\MS$ parameter $\hat s_Z^2$ is introduced:
\begin{equation} \label{eq400}
    \hat s_Z^2 = \hat s_{\text{ND}}^2 - \frac{\bar\alpha}{\pi} \LT( \frac16
    - \frac49 s^2 \RT) \ln \LT( \frac{m_t}{m_Z} \RT)^2
    = \hat s_{\text{ND}}^2 - 0.0002.
\end{equation}
From~\eqref{eq270} and~\eqref{eq400}, taken into account the latest
precision data fit value ${s_\Eff^\Lept}^2 = 0.2315 \pm 0.0002$, we get:
\begin{equation}
    \hat s_Z^2 = 0.2312 \pm 0.0002,
\end{equation}
where $\hat s_Z^2$ is uniquely defined both in the Standard Model and in its
extensions (unlike $\hat s^2$).

\nonumsection{Acknowledgments}

We are grateful to Z.~Berezhiani and A.~Rossi, discussion with whom
stimulated us to write this paper. We are also grateful to P.~Gambino for
very useful correspondence and for providing us with important references.

Investigations of M.~V. were supported by RFBR grants 96-02-18010,
96-15-96578, 98-02-17372, 98-02-17453 and INTAS-RFBR grant 95-05678.

\nonumsection{References}

\newpage

\end{document}